# ENERGY AWARE CLUSTERING PROTOCOL (EACP) FOR HETEROGENEOUS WSNS

Surender Kumar[1], Sumeet Kumar[2] and Bharat Bhushan[3]

[1]Department of Computer Engineering, Govt. Polytechnic, Ambala City, India
[2]Accenture Services Private Ltd., Bangalore, India
[3]Department of Computer Sc. & Applications, Khalsa College, Yamunanagar, India

## ABSTRACT

*Energy saving to prolong the network life is an important design issue while developing a new routing protocol for wireless sensor network. Clustering is a key technique for this and helps in maximizing the network lifetime and scalability. Most of the routing and data dissemination protocols of WSN assume a homogeneous network architecture, in which all sensors have the same capabilities in terms of battery power, communication, sensing, storage, and processing. Recently, there has been an interest in heterogeneous sensor networks, especially for real deployments. This research paper has proposed a new energy aware clustering protocol (EACP) for heterogeneous wireless sensor networks. Heterogeneity is introduced in EACP by using two types of nodes: normal and advanced. In EACP cluster heads for normal nodes are elected with the help of a probability scheme based on residual and average energy of the normal nodes. This will ensure that only the high residual normal nodes can become the cluster head in a round. Advanced nodes use a separate probability based scheme for cluster head election and they will further act as a gateway for normal cluster heads and transmit their data load to base station when they are not doing the duty of a cluster head. Finally a sleep state is suggested for some sensor nodes during cluster formation phase to save network energy. The performance of EACP is compared with SEP and simulation result shows the better result for stability period, network life and energy saving than SEP.*

## 1. INTRODUCTION

Wireless sensor network consists of a large number of tiny sensors and a sink or base station. Sink generally acts as a gateway for other network. Sensors not only sense the region, but also able to do computation, storage and communication to other sensors, including remote located base station by using a wireless medium. A tremendous amount of research activities has been going on in sensor networks due to their vital importance to a number of civilian and military applications. Wireless sensor networks are used for battlefield surveillance, for monitoring and tracking of nuclear plants and hazard locations [4],[5],[6],[7],[8],[9],[10]. Sensors are battery operated devices therefore it is crucial that they utilize their energy in an effective manner to increase the life of the network.

WSNs are usually deployed in a random way without any pre planning and exposed to terrible and dynamic environments. Sensor networks have to work for years without any attention from the external world. As a result, conventional algorithms which are suitable for other wireless networks like MANET (Mobile Adhoc Network) and cellular system cannot be applied directly to WSN [4], [5], [6]. Effective organization of the nodes to form clusters can save a significant amount of network energy [9]. In hierarchical or cluster based routing high energy nodes do the data processing and transmission while low energy node sense the region. It is an effective way to reduce the energy consumption of the network by performing data aggregation or data fusion to decrease the number of transmissions to base stations [4], [11], [21]. Sensor network are of two types: homogeneous and heterogeneous. When all the sensors of the network have the same capacity in terms of energy, storage and computation it is known as a homogeneous sensor network. However when the sensor nodes differ from each other in terms of energy, storage and computation it creates a heterogeneous network. Heterogeneity in sensor network helps in increasing the network lifetime and reliability. In this paper an energy aware clustering protocol is proposed to maximize the network lifetime. The protocol suggests a novel technique for cluster heads election and introduces a gateway concept for advance nodes to transfer the data load of normal cluster heads to base station. To save the energy of the network further, a sleep state is suggested for some sensor nodes.

The rest of this paper is organized as follows. In section 2, the related work is briefly reviewed. Section 3 explains the detail of purposed protocol. Section 4 exhibits the performance of EACP by simulations and compares it with SEP. The paper is concluded in section 5 with the direction for future work.

## 2. RELATED WORK

Hierarchical or cluster based routing helps in increasing energy efficiency, scalability, stability and network lifetime of the sensor network. Many cluster-based routing algorithms have been proposed for WSNs in the last few years. Low Energy Adaptive Clustering Hierarchy (LEACH) [1] is one of the pioneer protocols in this class and uses a single hop communication with an adaptive approach for cluster head election. There are two phases in LEACH: setup and steady state. In setup phase it elects the cluster head by generating a random number between 0 and 1 and then compares this number with a threshold value T (n), which is calculated by using Equation (1). If the generated number of a sensor node is less than this threshold, then it will become the cluster head for this round and advertises this by broadcasting a message in the network. A group of new CHs is elected during each round and in this way energy load is equally distributed among the sensor nodes.

$$T\,(n) \;=\; \left( \begin{array}{ll} \dfrac{p}{1 - p \times (r \bmod \dfrac{1}{p})} & \text{if n } \varepsilon \text{ G} \\[2em] 0 & \text{otherwise} \end{array} \right) \qquad (1)$$

Here p represents the percentage of nodes which can become the cluster head in a round, r denotes the current round and G represents the set of nodes which are not the cluster head in the last 1/p rounds.

In [1] an improvement over the LEACH protocol has given. This algorithm suggested a centralized approach for cluster head election, however steady phase is similar to LEACH. In LEACH there is no guarantee about the placement and/or number of cluster heads, but LEACH-C with the help of centralized algorithm can disperse CHs throughout the network.

One of the problems with LEACH is that it does not consider the remaining energy of a sensor node during the cluster head election. Due to this sometimes a node which does not have the sufficient energy can become a cluster head. To overcome this problem in [3] a new scheme for cluster head selection is proposed. When the remaining energy of a node is greater than 50% of the initial energy then LEACH algorithm threshold equation is applied (Equation 1), otherwise a new scheme which has the due consideration for the remaining energy of node is used (Equation 2).

$$T(n) = \begin{cases} \dfrac{p}{1 - p \times (r \bmod \dfrac{1}{p})} \times (2 \times p \times \dfrac{E_{residual}}{E_{init}}) & \text{if } n \, \varepsilon \, G \\ 0 & \text{otherwise} \end{cases} \qquad (2)$$

Here p denotes the percentage of nodes that can become the cluster head, $E_{residual}$ is the remaining energy of a node and $E_{init}$ is the initial energy of a node.

Power Efficient Gathering in Sensor Information System (PEGASIS) [12] is a greedy chain based algorithm for data gathering in wireless sensor network. Here each node forms a chain structure for transmitting the data load to base station. It achieves energy efficiency by sending data to only one of its neighbour node which fused the data with its own data and after that sends it to the next one hop neighbour. All the nodes are doing data fusion at its place so there is no rapid depletion of energy for base station nearer nodes.

EB - PEGASIS [13] is an energy efficient chaining algorithm in which a node will consider the average distance of formed chain. If the distance between closest node and its upstream node is longer than distance thresh (the distance thresh can be obtained from average distance of formed chain), the closest node is a "far node". If the closest node joins the chain, it will emerge as a "long chain". In this condition, the "far node" will search for a nearer node on formed chain. Through this method, EB-PEGASIS can avoid "long chain" effectively

Hybrid energy-efficient distributed (HEED) [14] algorithm is a distributed clustering algorithm for wireless sensor network. It chooses high residual energy nodes to become the cluster head and for load balancing re-clustering algorithm is executed periodically. In this way, nodes which have become cluster heads will have low probability to become cluster heads again. When the requirement is to distribute load among the cluster heads, node degree is used as a fitness function and the reciprocal of node degree when the dense clustering is desired.

Threshold sensitive Energy Efficient sensor Network protocol (TEEN) [15] is a protocol for time-critical applications where network operates in a reactive mode. TEEN uses a data-centric mechanism along with the hierarchical approach for its operation. The nodes which are close to each other, form clusters and this process continue to next level until the sink is reached. After the construction of clusters, the cluster head broadcasts hard and soft thresholds for sensing attributes to reduce the number of transmissions to base station. Disadvantage of TEEN is that it cannot be used when periodic reports are required because the user will not get any data if the thresholds have not reached. An extension of TEEN is Adaptive Threshold Sensitive Energy Efficient Sensor Network protocol (APTEEN) [16] which can capture data periodically and react to time critical events also.

Stable Election Protocol SEP [17] is an extension of the LEACH protocol for heterogeneous network. In SEP a small fraction of the nodes has more power than the normal nodes to create a heterogeneous network. For prolonging the stable region, SEP maintains energy consumption in a balanced manner.

DEEC [18] is an energy- efficient distributed clustering algorithm for heterogeneous wireless sensor networks. In DEEC, every sensor node independently elects itself as a cluster-head based on its residual energy and the average energy of the network. Density-Aware Energy-Efficient Clustering (DAEEC) [19] is a novel clustering algorithm proposed for uniformly distributed sensor networks to save energy and prolonging the network life. Energy-Balanced Routing Protocol for Data Gathering in Wireless Sensor Network (EBRP) [20] is a protocol which uses the physics potential concept for constructing a mixed virtual potential field and allows the moving of packets through the dense energy area towards the sink. In reference [22] an energy efficient clustering algorithm based on distance and residual energy is proposed which reduces the energy consumption and prolongs the network lifetime. Load-balanced clustering algorithm with distributed self organization for WSNs [23] combines the idea of distance and density distribution to propose an algorithm which forms more stable cluster structure and increases the network life. Multihop Energy Efficient Clustering & Data Aggregation Protocol for Heterogeneous WSNs (M-EECDA) [24] is a protocol for heterogeneous wireless sensor network. The protocol uses the idea of multihop communications and clustering to maximize the network lifetime.

## 3. ENERGY AWARE CLUSTERING PROTOCOL

In this section EACP (Energy Aware Clustering Protocol) is described for two level heterogeneous networks. The proposed protocol is the extension of SEP (Stable Election Protocol) [17] and has two types of the node: normal and advanced. The main goal of this protocol is to efficiently maintain the energy consumption and increases lifetime of the network. For implementing the protocol some reasonable assumptions have been made about the network and sensors which are as follows:

- Sensors are deployed randomly in a square region.
- The base station and sensors become stationary after deployment and the base station is located in middle of sensing region.
- Sensors are location unaware i.e. they do not have any information about their location.
- Sensors continuously sense the region and they always have data for sending to the base station.
- The battery of the sensors cannot be changed or charged as the nodes are densely deployed in a harsh environment.
- In the sensing region, there are two types of sensor nodes i.e. advanced and normal nodes. Advanced nodes have more energy than the normal nodes.

The model has n sensor nodes, which are deployed randomly in a $100 \times 100$ square meters region as displayed in Figure 1. Base station is located in middle of the sensing region and the distance of any node to its cluster head or sink is $\leq d_0$. The energy dissipated in the cluster head in a single round is given by Equation (3).

$$E_{CH} = \left(\frac{n}{k} - 1\right) \times L \cdot E_{elec} + \frac{n}{k} L \cdot E_{DA} + L \cdot E_{elec} + L \cdot \varepsilon_{fs} \cdot d_{toBS}^2 \qquad (3)$$

Where L is the no of bits of the message, $d_{toBS}$ is the average distance between the base station and cluster head and $E_{DA}$ is the energy required for performing data fusion or aggregation in a round. Since cluster members, send data to its cluster head therefore energy consumed in a non cluster head follows the free space path and it is given by Equation (4).

$$E_{NCH} = L \times (E_{elec} + \varepsilon_{fs} \times d_{toCH}^2) \qquad (4)$$

Where $d_{toCH}$ represents the average distance of the node from cluster head and the total energy consumed in a cluster is given by Equation (5).

$$E_{Cluster} = E_{CH} + \left(\frac{n}{k} - 1\right)E_{NCH} \approx E_{CH} + \left(\frac{n}{k}\right)E_{NCH} \tag{5}$$

The total dissipated energy of the network is given by Equations (6)

$$E_{round} = L \times \left(2nE_{elec} + nE_{DA} + k\,\varepsilon_{fs}d_{toBS}^2 + n\varepsilon_{fs}d_{toCH}^2\right) \tag{6}$$

The optimal number of clusters can be calculated by finding the derivative of $E_{round}$ with respect to k and equating it to zero.

$$k_{opt} = \frac{\sqrt{n}}{\sqrt{2\pi}} \sqrt{\frac{\varepsilon_{fs}}{\varepsilon_{mp}}} \frac{M}{d_{toBS}^2} \tag{7}$$

The average distance between cluster head and sink can be calculated [2], [17].

$$d_{toBS} = 0.765\frac{M}{2} \tag{8}$$

The optimal probability of a node to become the cluster head in a round is given by Equation 9.

$$p_{opt} = \frac{k_{opt}}{n} \tag{9}$$

## 3.1. Radio Dissipation Energy Model

Radio energy model as described in [1] is used for this protocol. Free space and multipath fading channel model both are considered here, depending upon the distance between the transmitter and receiver. When distance is less than a specific threshold value, then free space model used otherwise multipath loss model is considered. The amount of energy required to transmit $L$ bits packet over a distance, $d$ is given by Equation (10).

$$E_{Tx}(L,d) \quad \begin{array}{ll} L \times E_{elec} + L \times \varepsilon_{fs} \times d^2 & if\ (d < d_0) \\ L \times E_{elec} + L \times \varepsilon_{mp} \times d^4 & if\ (d \geq d_0) \end{array} \tag{10}$$

$E_{elec}$ is the electricity dissipated to run the transmitter or receiver circuitry. The parameters $\varepsilon_{mp}$ and $\varepsilon_{fs}$ is the amount of energy dissipated per bit in the radio frequency amplifier according to the distance $d_0$ which is given by the Equation (11).

$$d_0 = \sqrt{\frac{\epsilon_{fs}}{\epsilon_{mp}}} \tag{11}$$

The energy expended in receiving an $L$ bit message can be estimated by using Equation 12.

$$E_{Rx}(L) = L \times E_{elec} \tag{12}$$

## 3.2. Cluster Head Election Phase

EACP have two types of nodes (normal and advanced nodes). Let n be the total number of nodes and m be the fraction of n which are equipped with **α** times more energy than normal nodes. Powerful nodes are known as advanced nodes, and the rest (1 - m) × n as normal nodes. The initial energy of each normal node is $E_0$ and advanced node has $E_0 \times (1 + \alpha)$. Intuitively, advanced nodes have to become CHs more often than normal nodes because they have more energy than normal nodes. Network spatial density is not affected by this new setting. The value of $p_{opt}$ does not change but the total energy of the network is changed [17]. Heterogeneous network total initial energy is given by Equation 13.

$$E_{Total} = N \cdot (1 - m)E_0 + N \cdot mE_0(1 + \alpha) = N \cdot E_0(1 + \alpha m) \qquad (13)$$

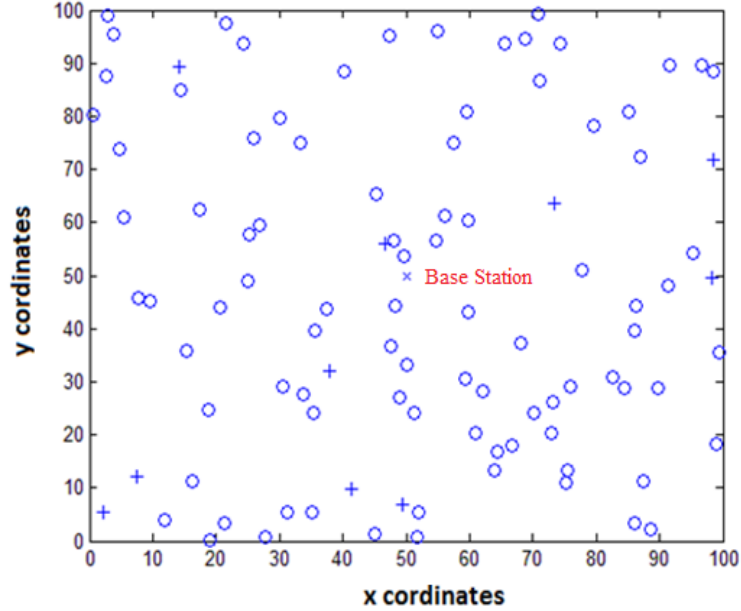

Figure 1 Sensing Region of EACP
(o - Normal, + - Advanced Node, x – Base Station)

Heterogeneity of nodes increases the system energy by $\alpha \cdot m$ times and for stable region optimization, new epoch must be equal to $\frac{1}{p_{opt}} \cdot (1 + \alpha \cdot m)$ [17]. Let $p_{nrm}$, $p_{adv}$ represents the weighted election probability of normal and advanced nodes respectively. Avg. number of cluster heads per round per epoch is $(n \times (1 + \alpha \cdot m) \cdot p_{nrm})$. The weighted probability for the various nodes can be found by using the Equation 14.

$$p_i = \begin{cases} \dfrac{p_{opt}}{(1 + \alpha \cdot m)} & if \ s_i \ is \ the \ normal \ node \\[4mm] \dfrac{p_{opt} \cdot (1 + \alpha)}{(1 + \alpha \cdot m)} & if \ si \ the \ advanced \ node \end{cases} \qquad (14)$$

In EACP threshold for cluster head election of normal and advanced nodes is determined by putting the above values in Equation (15)

$$T(s_i) = \begin{cases} \frac{p_i}{1-p_i\left(r \bmod \frac{i}{p_i}\right)} \times \frac{E(r)}{E(navg)} & if \ s_i \ \varepsilon \ G \\ \\ \frac{p_i}{1-p_i\left(r \bmod \frac{i}{p_i}\right)} & if \ s_i \ \varepsilon \ G' \\ \\ 0 & otherwise \end{cases} \qquad (15)$$

Where $G$ and $G'$ represents the set of normal or advanced nodes which are not elected as a cluster-head in the last $\frac{1}{p_i}$ rounds of the epoch, depending upon whether $s_i$ is a normal or advanced node. $E_{(r)}$ and $E_{(navg)}$ here denotes residual and average energy of normal node in a round. For calculating the average energy of normal nodes in a round EACP adopts the same procedure as used in DEEC [13]. Since the threshold calculation depends upon the average energy of normal sensor nodes in a round r, therefore it should be calculated. The average energy of normal nodes is estimated as:

$$E(navg) = \frac{1}{N} E_{ntotal}\left(1 - \frac{r}{R}\right) \qquad (16)$$

Here R represents the total round of the network lifetime and R can be estimated as

$$R = \frac{E_{ntotal}}{E_{round}} \qquad (17)$$

$E_{round}$ denotes the total energy dissipated in a round of the network and $E_{ntotal}$ is the total energy of normal nodes in the network. Cluster head threshold for normal nodes are multiplied by the ratio of residual energy and average energy of the normal nodes in a round because normal nodes have less energy as compared to advanced nodes therefore they should become cluster head only when they have sufficient energy for doing this duty.

## 3.3. Sleeping State

Most of the energy of the network is used in transmission therefore to reduce this energy EACP introduces three-tier architecture for normal sensor nodes. When a normal sensor node becomes cluster head in a round then after collecting the data from its members, it fuses the data and instead of sending the data to sink it will look for an advanced node such that

- Which is not a cluster head in this particular round?
- Distance of normal cluster head from advanced node is less than the distance between normal cluster head and base station.

If the normal cluster head succeed in this search then normal cluster head sends its data load to this advanced node for further sending to base station. If normal cluster head could not find any such advanced node then it sends the aggregated data to base station itself. Thus with the introduction of a gateway or three-tier architecture for normal cluster heads, EACP has reduced the transmission energy to prolong the stability period and lifetime of network.

Further after the election of cluster-heads in a round, non cluster head nodes try to join a nearby (considering the transmission power) cluster-head. However from an optimal energy saving point of view it is not a good choice in some scenarios. For example if a sensor node exists in the base station direction and distance of node from base station is less than distance of node from all nearby clusters (Figure 2). In Figure 2 node n1 has data load of L-bits for transmitting to base station. The nearest cluster-head to n1 is CH1. And, if node n1 joins to this cluster, it will spend the energy (18).

$$E1 = L \cdot E_{elec} + L \cdot \epsilon_{d2} \cdot d_2^x \tag{18}$$

Where
$$\begin{cases} x = 2, \epsilon\_d2 = 10 \text{ pj / bit / m}^2, & if \ d_2 < d_0 \\ x = 4, \epsilon_{d2} = 0.0013 \text{ pj / bit / m}^4, & if \ d_2 \geq d_0 \end{cases}$$

However if node n1 directly transfer data to the base station, this energy will be (19).

$$E2 = L \cdot E_{elec} + L \cdot \epsilon_{d3} \cdot d_3^y \tag{19}$$

Where
$$\begin{cases} y = 2, \epsilon_{d3} = 10 \text{ pj / bit / m}^2, & if \ d_3 < d_0 \\ y = 4, \epsilon_{d3} = 0.0013 \text{pj / bit / m}^4, & if \ d_3 \geq d_0 \end{cases}$$

Here positive coefficients $x$ and $y$ denote, energy dissipation radio model used. Obviously $E2 < E1$ but in this case an awful lot of redundant data will be collected at base station. To overcome this problem, EACP introduces a sleep state in the network.

When $E1 > E2$ then node $n1$ does not send the data to cluster head $CH1$,instead it enters into a sleep state and waits for the next round in which it either itself becomes a cluster head or finds a nearby cluster such that $E1 < E2$. It will remain in the sleep state for the maximum 4 rounds, if in these next 4 rounds it either becomes a cluster head or finds a nearby cluster such that $E1 < E2$ then it wakes up and performs its respective duty either of a cluster head or the members of a cluster head. If sensor node $n1$ is neither able to become the members of a cluster such that $E1 < E2$ nor become a cluster head in the next 4 rounds, then the sensor node wakes up and transmit the data load directly to the base station. The choice of maximum 4 rounds of sleep state is based on the fact that advanced nodes rotate the cluster head rotation cycle after every 4[th] round.

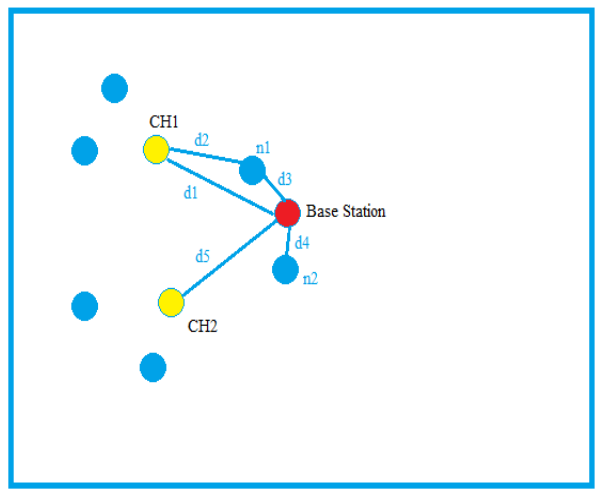

Figure 2 Cluster-head Selection for Transmission to Base Station

## 4. SIMULATIONS AND RESULTS

The performance of EACP is compared with SEP. For simulation 100 x 100 square meters region with 100 sensor nodes is used. Base station is located in middle of sensing region as shown in Figure 1. The normal nodes are represented by the symbol (o), advanced nodes (i.e. gateway nodes) with (+) and the base Station by (x). The various parameters of radio model used for the simulation are given in TABLE 1.

### 4.1. Performance Metrics

The performance metrics which are used for assessing the protocol are:

(i) **Stability Period:** It is defined by the time interval between network operation start until the death of the first sensor node.

(ii) **Network lifetime:** It is the time interval from the start of the network until the death of the last live node.

(iii) **Number of cluster heads per round:** This measures the number of nodes which would send the aggregated information to the base station.

(iv) **Number of Alive (total, advanced and normal) nodes per round:** This will measure the total number of live nodes of each type that has not yet expended all of their energy.

(v) **Throughput:** This measure the total rate of messages sent over the network, i.e. the rate of data sending from cluster heads to the sink as well as the rate of data sending from nodes to their respective cluster heads.

| Parameter | Value |
|-----------|-------|
| $E_{elec}$ | 5 nJ/bit |
| $\varepsilon_{fs}$ | 10 pJ/bit/m$^2$ |
| $\varepsilon_{mp}$ | 0.0013 pJ/bit/m$^4$ |
| $E_0$ | 0.5 J |
| $E_{DA}$ | 5 nJ/bit/message |
| Message Size | 4000 bits |
| $p_{opt}$ | 0.1 |
| $d_0$ | 70m |

Table 1 Radio Parameters of EACP

### 4.2. Performance Evaluation

The performance of EACP is evaluated by introducing various parameters of heterogeneity in the system and the following two cases of heterogeneity have been considered for this.

**Case 1: m = 0.1, a =5**

In this case there are **10** advanced nodes with **5** times more energy than normal nodes and **90** normal nodes are deployed randomly in the sensing region. Figure 3 shows that the network lifetime of EACP is more than SEP as last node dies in EACP at **9072$^{th}$** round and in SEP it dies at **6000$^{th}$** round i.e. network life has been increased by approximately **51%.** In SEP first node dies at **1255$^{th}$** round and in EACP it happens at **1479$^{th}$** round. It means that EACP has extended the stability period of the system by **18%.** Figure 4 plots the number of alive nodes per round and they are more in EACP as compared to SEP. Throughput of the network is plotted in Figure 5 i.e. total number of messages sent to the base station and from the graph it is clear that EACP has sent more messages to the base station than SEP. Figure 6 plots the total residual energy (in joules) of the network per round in EACP and SEP and it clearly shows the advantages of EACP over SEP. Initially up to **500$^{th}$** round total remaining energy in SEP and EACP is almost equal, but after that energy in SEP gradually depleting faster than EACP and entire energy has been depleted at **6000$^{th}$** round however in EACP it occurs at **9072$^{th}$** round.

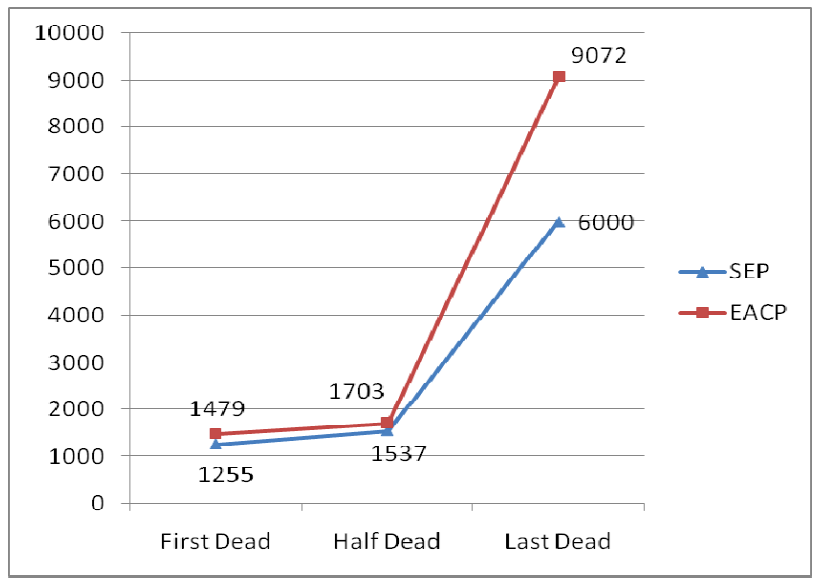

Figure 3 Rounds for First, Half and Last Node Death in EACP & SEP

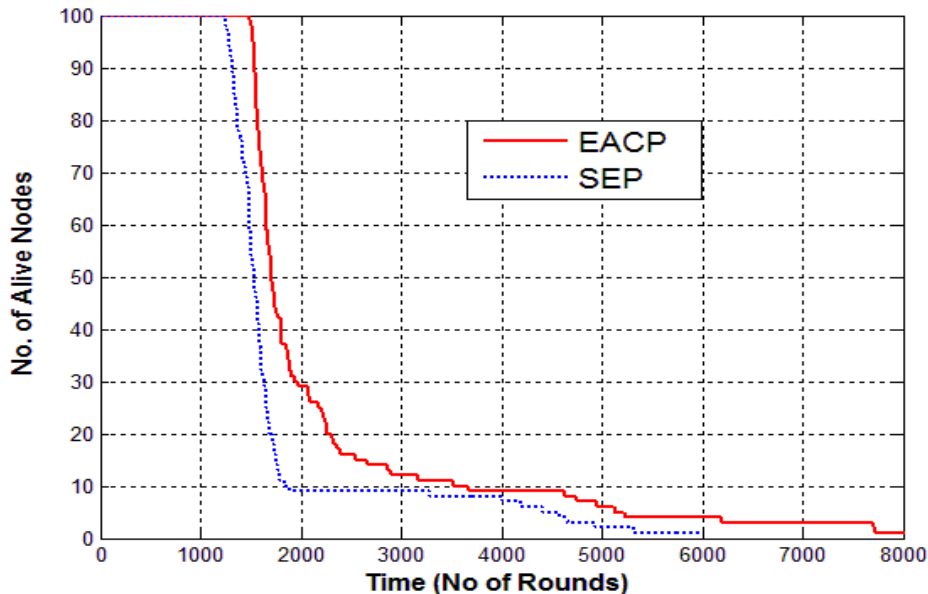

Figure 4 Comparisons of No. of Alive Nodes per Round in EACP & SEP

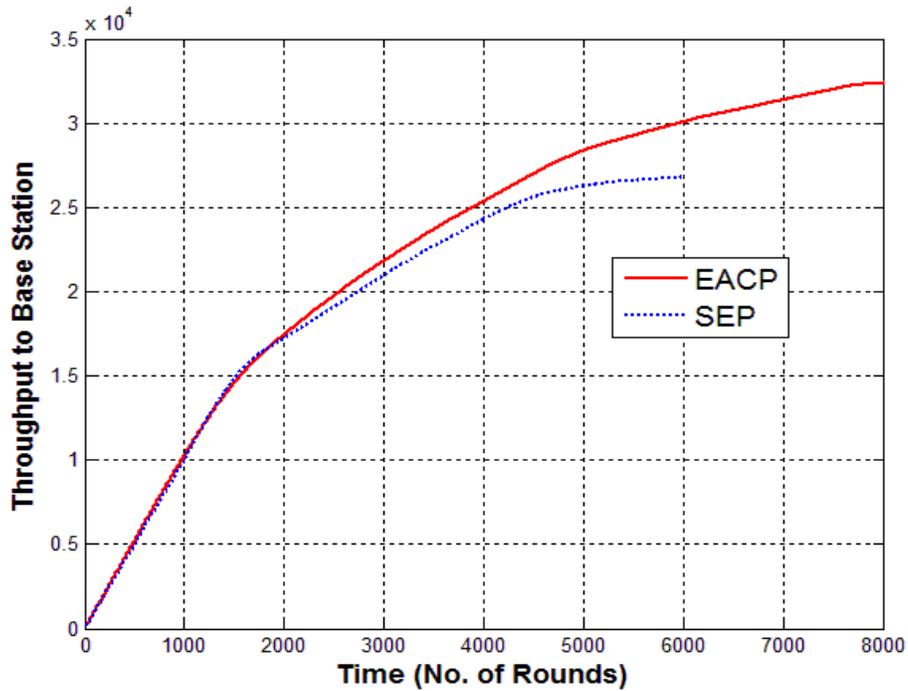

Figure 5 Throughput Comparisons in EACP & SEP

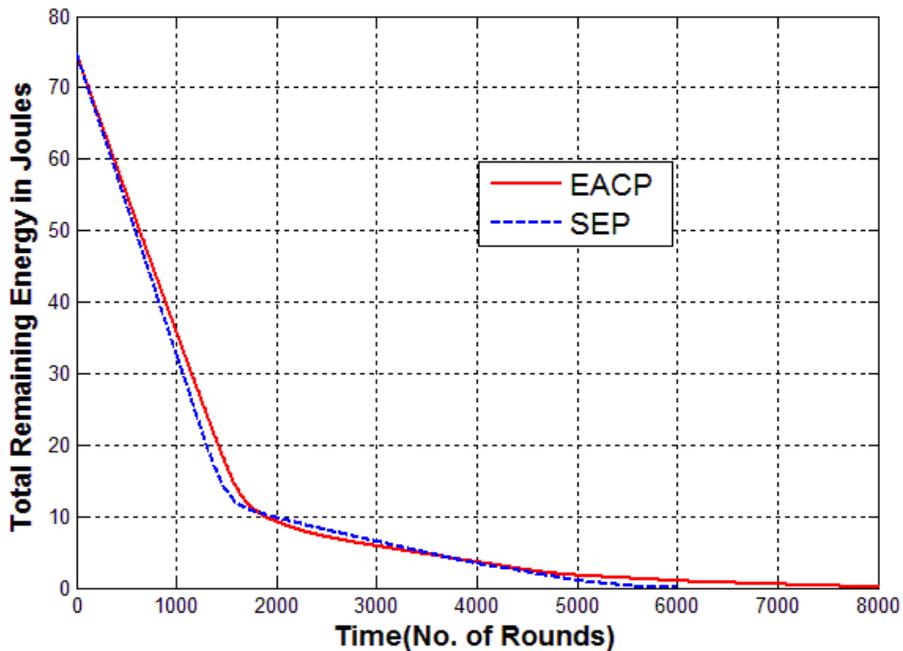

Figure 6 Total Remaining Energy Comparisons in EACP & SEP

**Case 2: m = 0.2, a = 3**

In this case there are **20** advanced nodes having **3** times more energy than normal nodes and **80** normal sensor nodes, are randomly dispersed in the sensing region. SEP has lower network life than EACP because last node in SEP dies at **6451**th round and in EACP it occurs at **8807**th round. Thus EACP extends the network lifetime approximately by **37%**. Stability period of EACP is also more as compared to SEP because first node dies in EACP at **1505**[th] round and in SEP it happens at **1335**[th] round. It means that stability period of the system is approximately **13%** more in EACP. Figure 8 shows that no of alive nodes per round in EACP and SEP and it clearly indicates the edge of EACP over SEP. Figure 9 shows the throughput i.e. total numbers of messages sent to the base station and they are also more in EACP. Figure 10 plots the total remaining energy (in joules) per round and from the figure it is clear that EACP has more remaining energy per round than SEP.

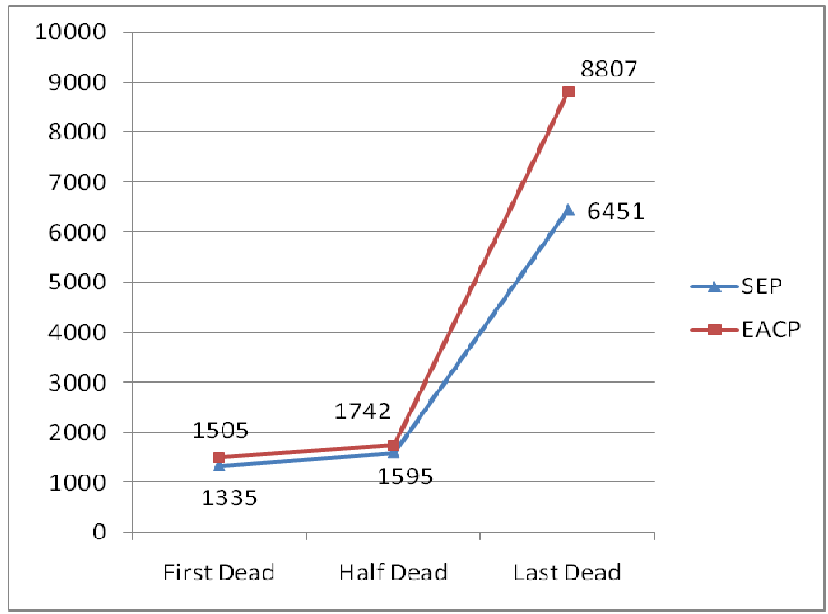

Figure 7 Rounds for First, Half and Last Death in EACP & SEP

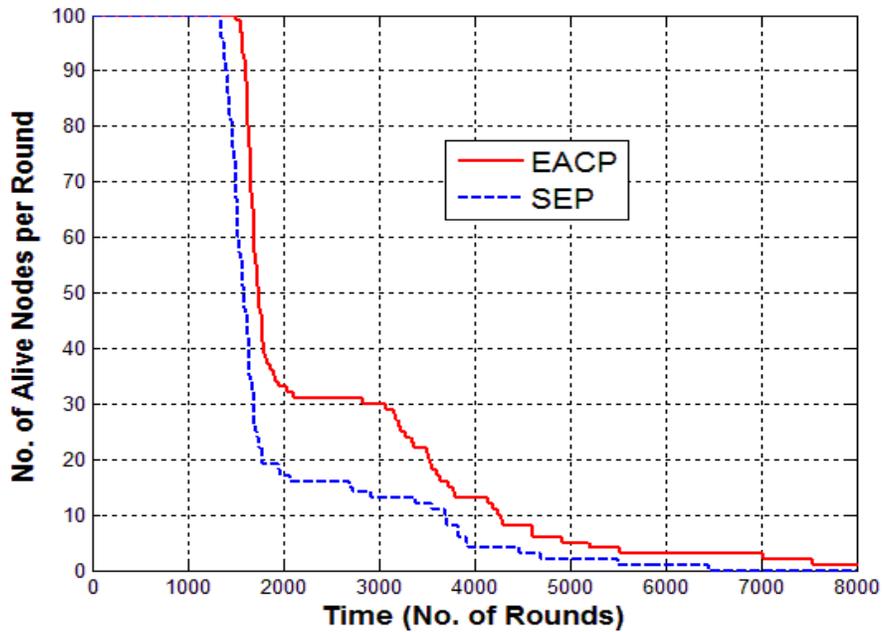

Figure 8 Comparisons of Alive Nodes per Round in EACP & SEP

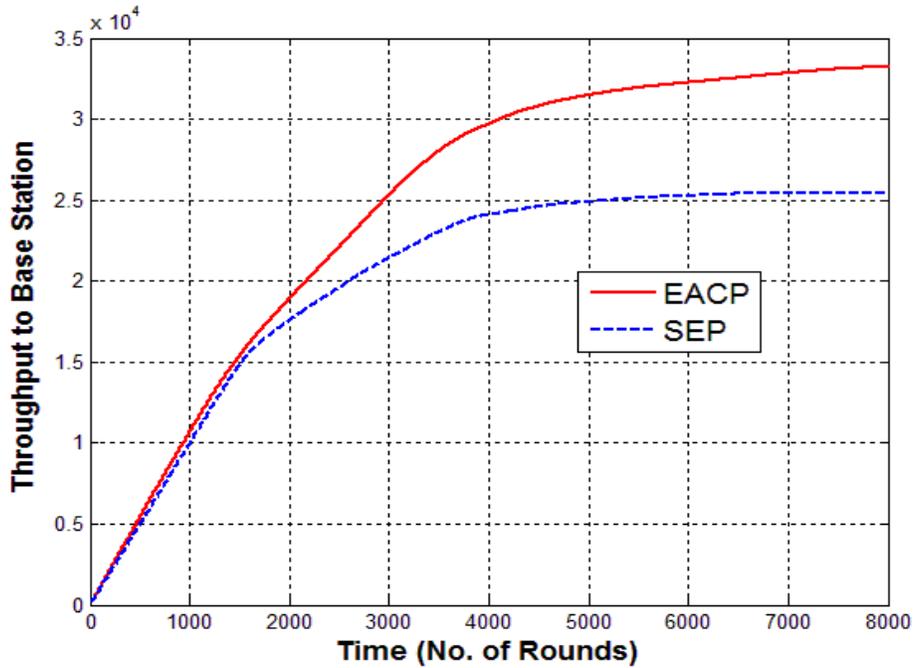

Figure 9 Throughput Comparisons in EACP & SEP

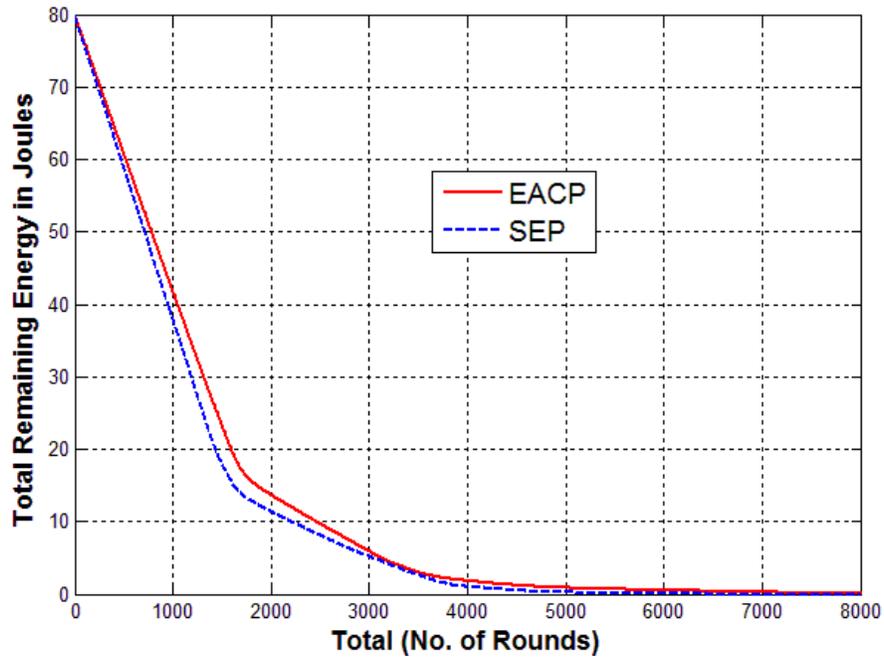

Figure 10 Comparisons of Total Remaining Energy in EACP & SEP

## 5. CONCLUSIONS

In this paper EACP (Energy Aware Clustering Protocol) for two level heterogeneous sensor networks has been proposed. The proposed protocol takes the full advantage of heterogeneity and improves the lifetime, stable region and throughput of the network. Most of the network energy is depleted in transmission therefore EACP suggests an energy saving scheme for normal nodes during cluster head election phase. Normal nodes can become a cluster head only when they have enough energy to do this duty. For this threshold equation of normal nodes is modified by multiplying it with the ratio of residual energy and total average energy of normal nodes in a round. This will ensure that only high residual energy normal nodes can become cluster head in a round. EACP has introduced three-tier architecture for normal cluster heads and their data loads are taken over by advance nodes when they are not performing the duty of a cluster head in any specific round. When the nearby cluster head joining is not an energy saving option for non cluster head nodes, then protocol suggests a sleep state for them so that the energy of the network can be saved. Through simulation, it has been found that EACP extends the stable region, network life and throughput of the system significantly. Future directions of this research will study using some other techniques like fuzzy logic, genetic algorithm etc. for cluster head election.

**Surender Kumar** received his Master of Computer Applications degree from Kurukshetra University, Kurukshetra, India and M.Phil (Computer Science) from Madurai Kamraj University, Madurai, India. He is currently working as System Programmer at Govt. Polytechnic, Ambala City. He has more than 13 years of experience and his research interests include Wireless Sensor network, Computer Network, Software Engineering and Network Security.

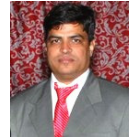

**Sumeet Kumar** is currently working with Accenture Services Private Limited, Bangalore, India as Software tester. He passed his Bachelor of Technology in Computer Science & Engineering from Kurukshetra University, Kurukshetra. He has expertise in domain banking software testing in IVR based applications, mainframes and Web Applications. He has also worked with Panoply Soft Private Ltd. India, Chandigarh as UI tester and web application developer. His areas of interests include Mobile Adhoc Networks, Android Applications, Web Mining and Software Engineering.

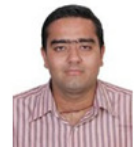

**Dr. Bharat Bhushan** received his PhD from Kurukshetra University, India. He is currently working as Head, Department of Computer Science & Applications, Guru Nanak Khalsa College, Yamunanagar, India. He has more than 29 years of experience and published many research papers in journals and conferences at National and International level. His research interests include Software Engineering, Digital Electronics, Networking and Simulation Experiments.

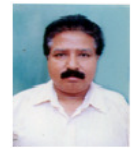